# Unidirectional Imaging using Deep Learning-Designed Materials

Jingxi Li[1,2,3], Tianyi Gan[1,3], Yifan Zhao[1,3], Bijie Bai[1,2,3], Che-Yung Shen[1,2,3], Songyu Sun[1], Mona Jarrahi[1,3], and Aydogan Ozcan[1,2,3*]

[1]Electrical and Computer Engineering Department, University of California, Los Angeles, CA, 90095, USA

[2]Bioengineering Department, University of California, Los Angeles, CA, 90095, USA

[3]California NanoSystems Institute (CNSI), University of California, Los Angeles, CA, 90095, USA

*Correspondence to: ozcan@ucla.edu

## Abstract

A unidirectional imager would only permit image formation along one direction, from an input field-of-view (FOV) A to an output FOV B, and in the reverse path, B→A, the image formation would be blocked. Here, we report the first demonstration of unidirectional imagers, presenting polarization-insensitive and broadband unidirectional imaging based on successive diffractive layers that are linear and isotropic. These diffractive layers are optimized using deep learning and consist of hundreds of thousands of diffractive phase features, which collectively modulate the incoming fields and project an intensity image of the input onto an output FOV (i.e., A→B), while blocking the image formation in the reverse direction, B→A. After their deep learning-based training, the resulting diffractive layers are fabricated to form a unidirectional imager. As a reciprocal device, the diffractive unidirectional imager has asymmetric mode processing capabilities in the forward and backward directions, where the optical modes from B→A are selectively guided/scattered to miss the output FOV, whereas for the forward direction (A→B) such modal losses are minimized, yielding an ideal imaging system between the input and output FOVs. Although trained using monochromatic illumination, the diffractive unidirectional imager maintains its functionality over a large spectral band and works under broadband illumination. We experimentally validated this unidirectional imager using terahertz radiation, very well matching our numerical results. Using the same deep learning-based design strategy, we also created a wavelength-selective unidirectional imager, where two unidirectional imaging operations, in reverse directions, are multiplexed through different illumination wavelengths. Diffractive unidirectional imaging using structured materials will have numerous applications in e.g., security, defense, telecommunications and privacy protection.

# Introduction

Optical imaging applications have permeated every corner of modern industry and daily life. A myriad of optical imaging methods have flourished along with the progress of physics and information technologies, resulting in imaging systems such as super-resolution microscopes[1,2], space telescopes[3–5] and ultrafast cameras[6,7] that cover various spatial and temporal scales at different bands of the electromagnetic spectrum. With the recent rise of machine learning technologies, researchers have also started using deep learning algorithms to design optical imaging devices based on massive image data and graphics processing units (GPUs), achieving optical imaging designs that, in some cases, surpass what can be obtained through physical intuition and engineering experience[8–14].

Standard optical imaging systems composed of linear and time-invariant components are reciprocal, and the image formation process is maintained after swapping the positions of the input and output fields-of-view (FOVs). If one could introduce a unidirectional imager, the imaging black-box would project an image of an input object FOV (A) onto an output FOV (B) through the forward path (A→B), whereas the backward path (B→A) would block the image formation process (see Fig. 1a).

To design a unidirectional imager, one general approach would be to break electromagnetic reciprocity: one can use e.g., magneto-optic effect (the Faraday effect)[15–17], temporal modulation of the electromagnetic medium[18,19], or other nonlinear optical effects[20–27]. However, realizing such non-reciprocal systems for unidirectional imaging over a sample FOV with many pixels poses challenges due to high fabrication costs, bulky and complicated set-ups/materials and/or high-power illumination light sources. Alternative approaches have also been used to achieve unidirectional optical transmission from one point to another without using optical isolators. One of the most common practices is using a quarter-wave plate and a polarization beam splitter; this approach for point-to-point transmission is polarization sensitive and results in an output with only circular polarization. Other approaches include using asymmetric isotropic dielectric gratings[28–31] and double-layered metamaterials[32] to create different spatial mode transmission properties along the two directions. However, these methods are designed for relatively simple input modes and face challenges in off-axis directions, thus making them difficult to form imaging systems even with relatively low numerical apertures.

Despite all the advances in materials science and engineering and optical system design, there is no unidirectional imaging system reported to date, where the forward imaging process (A→B) is permitted and the reverse imaging path (B→A) is all-optically blocked.

Here, we report the first demonstration of unidirectional imagers and design polarization-insensitive and broadband unidirectional imaging systems based on isotropic structured linear materials (see Figs. 1b and c). Without using any lenses commonly employed in imaging, here we optimize a set of successive dielectric diffractive layers consisting of hundreds of thousands of diffractive features with learnable thickness (phase) values that collectively modulate the incoming optical fields from an input FOV. After being trained using deep learning,[33–46] the resulting diffractive layers are physically fabricated to form a unidirectional imager, which performs polarization-insensitive imaging of the input FOV with high structural fidelity and power efficiency in the forward direction (A→B), while blocking the image transmission in the backward direction, not only penalizing the diffraction efficiency from B→A but also losing the structural similarity or resemblance to the input images. Despite being trained using only MNIST handwritten digits, these diffractive unidirectional imagers are able to generalize to more complicated input images from other datasets, demonstrating their external generalization capability and serving as a general-purpose unidirectional imager from A→B. Although these diffractive unidirectional imagers were trained using monochromatic illumination at a wavelength of λ, they maintain unidirectional imaging functionality under broadband illumination, over a large spectral band that uniformly covers e.g., 0.85×λ to 1.15×λ.

We experimentally confirmed the success of this unidirectional imaging concept using terahertz waves and a 3D-printed diffractive imager and revealed a very good agreement with our numerical results by providing clear and intense images of the input objects in the forward direction and blocking the image formation process in the backward direction. Using the same deep learning-based training strategy, we also designed a wavelength-selective unidirectional imager that performs unidirectional imaging along one direction (A→B) at a predetermined wavelength and along the opposite direction (B→A) at another predetermined wavelength. With this wavelength-multiplexed unidirectional imaging design, the operation direction of the diffractive unidirectional imager can be switched (back and forth) based on the illumination wavelength, improving the versatility and flexibility of the imaging system.

The optical designs of these diffractive unidirectional imagers have a compact size, axially spanning ~80-100λ. Such a thin footprint would allow these unidirectional imagers to be integrated into existing optical systems that operate at various scales and wavelengths. While we considered here spatially coherent illumination, the same design framework and diffractive feature optimization method can also be applied to spatially incoherent scenes. Polarization-insensitive and broadband unidirectional imaging using linear and isotropic structured materials will find various applications in security, defense, privacy protection and telecommunications among others.

## Results

**Diffractive unidirectional imager using reciprocal structured materials**

Figure 1a depicts the general concept of unidirectional imaging. To create a unidirectional imager using reciprocal structured materials that are linear and isotropic, we optimized the structure of phase-only diffractive layers (i.e., $L_1$, $L_2$, …, $L_5$), as illustrated in Figs. 1b and c. In our design, all the diffractive layers share the same number of diffractive phase features (200 × 200), where each dielectric feature has a lateral size of ~λ/2 and a trainable/learnable thickness providing a phase modulation range of 0-2π. The diffractive layers are connected to each other and the input/output FOVs through free space (air), resulting in a compact system with a total length of 80λ (see Fig. 2a). The thickness profiles of these diffractive layers were iteratively updated in a data-driven fashion using 55,000 distinct images of the MNIST handwritten digits (see the Methods section). A custom loss function is used to simultaneously achieve the following three objectives: (1) minimize the structural differences between the forward output images (A→B) and the ground truth images based on the normalized mean square error (MSE), (2) maximize the output diffraction efficiency (overall transmission) in the forward path, A→B, and (3) minimize the output diffraction efficiency in the backward path, B→A. More information about the architecture of the diffractive unidirectional imager, loss functions and other training-related implementation details can be found in the Methods section.

The resulting diffractive layers after the completion of the training are shown in Fig. 2c, revealing their phase modulation coefficients. This diffractive unidirectional imager design was numerically tested using the MNIST test dataset, which consists of 10,000 handwritten digit images that were never seen by the diffractive model during the training process. We report some of these blind testing results in Fig. 2d for both the forward and backward directions, clearly illustrating the internal generalization of the resulting diffractive imager to new, unseen input images from the same dataset. We also quantified the performance of this diffractive unidirectional imager for both the forward and backward directions based on the following metrics: (1) the normalized MSE and (2) the Pearson Correlation Coefficients (PCC) between the input and output images (denoted as "output MSE" and "output PCC"), and (3) the output diffraction efficiencies; these metrics were calculated using the same set of MNIST test images, never seen before. As shown in Figs. 3a-b, the forward (A→B) and backward (B→A) paths of the diffractive unidirectional imager shown in Fig. 2c provide output MSE values of $(5.68\pm1.56)\times10^{-5}$ and $(0.919\pm0.048)\times10^{-3}$, respectively, and their output PCC values are calculated as $0.9740 \pm 0.0065$ and $0.3839 \pm 0.0685$, respectively. A similar asymmetric behavior between the forward and backward imaging directions is also observed for the output diffraction efficiency metric as shown in Fig. 3c: the output diffraction efficiency of A→B is found as $93.50\% \pm 1.56\%$ whereas it is reduced to $1.57\% \pm 0.44\%$ for B→A, which constitutes an average image power suppression ratio of ~60-fold in the reverse direction compared to the forward imaging direction. Equally important as this poor diffraction efficiency for B→A is the fact that the weak optical field in the reverse direction does not have spatial resemblance to the input objects as revealed by a poor average PCC value of ~0.38 for B→A. These results demonstrate and quantify the internal generalization success of our diffractive unidirectional imager: the input images can be successfully imaged with high structural fidelity and power efficiency along the forward direction of the diffractive imager, while the backward imaging operation B→A is blocked by substantially reducing the output diffraction efficiency and distorting the structural resemblance between the input and output images.

To better understand the working principles of this diffractive unidirectional imager, next we consider the three-dimensional space formed by all the diffractive layers and the input/output planes as a diffractive volume, and categorize/group the optical fields propagating within this volume as part of different spatial modes: (1) the optical modes that finally arrive at the target output FOV, i.e., at FOV B for A→B and at FOV A for B→A; (2) the optical modes arriving at the output plane but outside the target output FOV; and (3) the unbounded optical modes that do not reach the output planes; since the diffractive layers are axially separated by $>10\lambda$, there are no evanescent waves being considered here. We calculated the power distribution percentages of each one of these types of optical modes for both A→B and B→A for each test image, and reported their average values across the 10,000 test images in Fig. 3d (see the Methods section for details). The results summarized in Fig. 3d clearly reveal that, in the forward path (A→B) of the diffractive unidirectional imager, the majority of the input power (>93.5%) is coupled to the imaging modes that arrive at the output FOV B, forming high-quality images of the input objects with a mean PCC of 0.974, while the optical modes that fall outside the FOV B and the unbound modes are minimal, accounting for only ~2.95% and ~3.54% of the input total power, respectively. In contrast, the backward imaging path (B→A) of the same diffractive unidirectional imager steers most of the input power into the non-imaging modes that fall outside the FOV A or escape out of the diffractive volume through the unbounded modes, which correspond to power percentages of ~34.8% and ~63.6%, respectively. For B→A the optical modes that arrive at the FOV A only constitute on average ~1.57% of the input total power; however, these optical modes are not only weak, but they are also significantly aberrated by the diffractive unidirectional imager, resulting in very poor output images, with a mean PCC value of ~0.38.

The underlying reason for these contrasting power distributions in the two imaging directions stems from the different order of the diffractive layers as the light passes through them: in the forward operation A→B, the diffractive layers arranged in the order of $L_1$ to $L_5$ manage to focus most of the light waves within the output FOV B forming high-quality images, while in the backward operation B→A, the same set of diffractive layers arranged in the reversed order (i.e., $L_5$ to $L_1$) couple most of the transmitted input optical fields into the non-imaging modes that never arrive at FOV A, while at the same time scrambling/messing the distributions of the imaging modes that arrive at FOV A, forming significantly weak and poor quality images. It is worth noting that, since the presented diffractive unidirectional imager is composed of linear, time-invariant and isotropic materials, it forms a reciprocal system that is polarization insensitive. In experimental implementations (reported below) due to absorption related losses a diffractive unidirectional imager also exhibits time-reversal asymmetry.

To further highlight the capabilities of our diffractive unidirectional imager (which was trained using handwritten digits), we also tested its external generalization using other datasets: the EMNIST dataset which contains images of handwritten English letters, and the Fashion-MNIST dataset which contains images of various fashion products. The blind testing results on these two new datasets using the diffractive unidirectional imager of Fig. 2c are exemplified in Fig. 2e, which once again confirm its success. We also quantified the imaging resolution performance of this diffractive unidirectional imager using gratings as resolution test targets, which were also never used in the training phase (see Fig. 4). Our results reveal that the diffractive unidirectional imager can resolve a minimum linewidth of ~4λ in the forward path A→B, while blocking the image formation in the reverse path B→A, as expected. These results once again prove that the training of the diffractive unidirectional imager is successful in approximating a general-purpose imaging operation in the forward path, although we only used handwritten digits during its training.

**Spectral response of the diffractive unidirectional imager**

Next, we explored the spectral response of the diffractive unidirectional imager reported in Fig. 2 under different illumination wavelengths that deviate from the training illumination wavelength ($\lambda_{train}=\lambda$). The results of this analysis are reported in Fig. 5, where the output image PCC and diffraction efficiency values of the diffractive unidirectional imager of Fig. 2 were tested as a function of the illumination wavelength. Although this diffractive unidirectional imager was only trained at a single illumination wavelength ($\lambda$), it also works well over a large spectral range as shown in Fig. 5. Our results reveal that the imaging performance in the forward path (A→B) remains very good with an output image PCC value of ≥ 0.85 and an output diffraction efficiency of ≥ 85.5% within the entire spectral range [$\lambda_L : \lambda_R$], where $\lambda_L = 0.92 \times \lambda$ and $\lambda_R = 1.11 \times \lambda$ (see Figs. 5a-b). Within the same spectral range defined by [$\lambda_L : \lambda_R$], the power suppression ratio between the forward and backward imaging paths always remains ≥ 17.4× and the output diffraction efficiency of the reverse path (B→A) remains ≤ 5.49% (see Fig. 5b), indicating the success of the diffractive unidirectional imager over a large spectral band, despite the fact that it was only trained with monochromatic illumination at λ. Figure 5d further reports examples of test objects (never seen during the training) that are simultaneously illuminated by a continuum of wavelengths, covering two different broadband illumination cases: (i) [0.92×λ : 1.11×λ] and (ii) [0.85×λ :

1.15×λ]. The forward and backward imaging results for these two broadband illumination cases shown in Fig. 5d clearly illustrate the success of the diffractive unidirectional imager under broadband illumination. We should emphasize that these broadband unidirectional imaging results can be further enhanced by training the diffractive layers using a set of wavelengths sampled from a desired spectral band, as an alternative to using a single training wavelength.

**Experimental validation of the diffractive unidirectional imager design**

We experimentally validated our diffractive unidirectional imager using a monochromatic continuous-wave terahertz (THz) illumination at λ = 0.75 mm, as shown in Fig. 6a. A schematic diagram of the THz set-up is shown in Fig. 6b, and its implementation details are reported in the Methods section. For this experimental validation, we designed a diffractive unidirectional imager composed of three diffractive layers, where each layer contains 100 × 100 learnable diffractive features, each with a lateral size of 0.64λ (dictated by the resolution of our 3D-printer). The axial spacing between any two adjacent layers (including the diffractive layers and the input/output planes) is chosen as ~26.7λ. Different from earlier designs, here we also took into account the material absorption using the complex-valued refractive index of the diffractive material in our optical model, such that the optical fields absorbed by the diffractive layers are also considered in our design (which will be referred to as the "absorbed modes" in the following discussion). Moreover, to overcome the undesired performance degradation that may be caused by the misalignment errors in an imperfect physical assembly of the diffractive layers, we also adopted a "vaccination" strategy in our design by introducing random displacements applied to the diffractive layers during the training process, which enabled the final converged diffractive unidirectional imager to become more resilient to potential misalignment errors (see the Methods section). After the training is complete, the resulting diffractive layers were fabricated using a 3D-printer (Figs. 6c-d). Also see Supplementary Fig. S1 for the numerical performance analysis of this converged diffractive design using blind testing objects.

In our experiments, we tested the performance of this 3D-fabricated diffractive unidirectional imager along the forward and backward directions, as illustrated in Figs. 7a-b. 10 different handwritten digit samples from the blind testing set (never used in the training) were used as the input test objects, also 3D-printed. These experimental imaging results for A→B and B→A are shown in Fig. 7c, which present a good agreement with their numerical simulated counterparts, very well matching the input images. As expected, 3D-printed diffractive unidirectional imager faithfully imaged the input objects in its forward direction and successfully blocked the image formation in the backward direction; these results constitute the first demonstration of unidirectional imaging.

**Wavelength-multiplexed unidirectional diffractive imagers**

Next, we consider a more challenging task: combining two diffractive unidirectional imagers that operate in opposite directions, where the direction of imaging is controlled by the illumination wavelength. The resulting diffractive system forms a novel wavelength-multiplexed unidirectional imager, where the image formation from A→B and B→A is maintained at $\lambda_1$ and $\lambda_2$ illumination wavelengths, respectively, whereas the image formation from B→A and A→B is blocked at $\lambda_1$ and $\lambda_2$, respectively (see Figs. 8-9). To implement this wavelength-multiplexed unidirectional imaging concept, we designed a new diffractive imager that operates at $\lambda_1$ and $\lambda_2 = 1.13 \times \lambda_1$ wavelengths and used an additional penalty term in the training loss function to improve the performance of the image blocking operations in each direction, A→B and B→A. More details about the numerical modeling and the training loss function for this wavelength-multiplexed diffractive design can be found in the Methods section.

We trained this wavelength-multiplexed unidirectional diffractive imager using handwritten digit images as before; the resulting, optimized diffractive layers are reported in Supplementary Fig. S2. Following its training, the diffractive imager was blindly tested using 10,000 MNIST test images that were never used during the training process, with some representative testing results presented in Fig. 9c. These results indicate that the wavelength-multiplexed diffractive unidirectional imager successfully performs two separate unidirectional imaging operations, in reverse directions, the behavior of which is controlled by the illumination wavelength; at $\lambda_1$ A→B image formation is permitted and B→A is blocked, whereas at $\lambda_2$ B→A image formation is permitted and A→B is blocked.

We also analyzed the imaging performance of this wavelength-multiplexed unidirectional diffractive imager as shown in Figs. 10a-c. At the first wavelength channel $\lambda_1$, the output PCC values for the forward (A→B) and backward (B→A) directions are calculated as 0.9428 ± 0.0154 and 0.1228 ± 0.0985, respectively, revealing an excellent image quality contrast between the two directions (see Fig. 10b). Similarly, the output diffraction efficiencies for the forward and backward directions at $\lambda_1$ are quantified as 65.82% ± 3.57% and 3.62% ± 0.72%, respectively (Fig. 10c). In contrast, the second wavelength channel $\lambda_2$ of this diffractive model performs unidirectional imaging along the direction opposite to that of the first wavelength, providing output PCC values of 0.9378 ± 0.0187 (B→A) and 0.0840 ± 0.0739 (A→B); see Fig. 10b. Similarly, the output diffraction efficiencies at $\lambda_2$ were quantified as 51.81% ± 3.77% (B→A) and 2.57% ± 0.36% (A→B). These findings can be further understood by investigating the power distribution within this wavelength-multiplexed unidirectional diffractive imager, which is reported in Fig. 10d. This power distribution analysis within the diffractive volume clearly shows how two different wavelengths ($\lambda_1$ and $\lambda_2$) along the same spatial direction (e.g., A→B) can result in very different distributions of spatial modes, performing unidirectional imaging in opposite directions, following the same physical behavior reported in Fig. 3d, except that this time it is wavelength-multiplexed, controlling the direction of imaging. Such an exotic wavelength-multiplexed unidirectional imaging system cannot be achieved using simple spectral filters such as absorption or thin-film filters since the use of a spectral filter at one wavelength channel (for example to block A→B at $\lambda_2$) would immediately also block the reverse direction (B→A at $\lambda_2$), violating the desired goal.

We should also note that since this wavelength-multiplexed unidirectional imager was trained at two distinct wavelengths that control the opposite directions of imaging, the spectral response of the resulting diffractive imager, after its optimization, is vastly different from the broadband response of the earlier designs, reported in e.g. Fig. 5. Supplementary Fig. S3 reveals that the wavelength-multiplexed unidirectional imager (as desired and expected) switches its spectral behavior in the range between $\lambda_1$ and $\lambda_2$ since its training aimed unidirectional imaging at opposite directions at these two predetermined wavelengths. Therefore, this spectral response that is summarized in Supplementary Fig. S3 is in line with the training goals of this wavelength-multiplexed unidirectional imager. However, it still maintains its unidirectional imaging capability over a range of wavelengths in both directions. For example, Supplementary Fig. S3 reveals that the output image PCC values for A→B remain ≥ 0.85 within the entire spectral range covered by 0.975×$\lambda_1$ to 1.022×$\lambda_1$ without any considerable increase in the diffraction efficiency for the reverse path, B→A. Similarly, the output image PCC values for B→A remain ≥ 0.85 within the entire spectral range covered by 0.968×$λ_2$ to 1.029×$λ_2$ without any noticeable increase in the diffraction efficiency for the reverse path, A→B, within the same spectral band. These results highlighted in Supplementary Fig. S3 indicate that the wavelength-multiplexed unidirectional imager can also operate over a continuum of wavelengths around $λ_1$ (A→B) and $λ_2$ (B→A), although the width of these bands are narrower compared to the broadband imaging results reported in Fig. 5.

Finally, we also tested the external generalization capability of this wavelength-multiplexed unidirectional imager on different datasets: handwritten letter images and fashion products. The corresponding imaging results are shown in Fig. 9d, once again confirming that our diffractive model successfully converged to a data-independent, generic imager where unidirectional imaging of various input objects can be achieved along either the forward or backward directions that can be switched/controlled by the illumination wavelength.

## Discussion

Our results constitute the first demonstration of unidirectional imaging. This framework utilizes structured materials formed by phase-only diffractive layers optimized through deep learning, and does not rely on non-reciprocal components, nonlinear materials or an external magnetic field bias. Due to the use of isotropic diffractive materials, the operation of our unidirectional imager is insensitive to the polarization of the input light, also preserving the input polarization state at the output. Furthermore, the presented diffractive unidirectional imagers maintain unidirectional imaging functionality under broadband illumination, over a large spectral band that covers e.g., 0.85×λ to 1.15×λ, despite the fact that they were only trained using monochromatic illumination at λ. This broadband imaging performance can be further enhanced, covering even larger input bandwidths, by training the diffractive layers of the unidirectional imager using a set of illumination wavelengths randomly sampled from the desired spectral band of operation.

Throughout this manuscript, we presented diffractive unidirectional imagers with input and output FOVs that have 28×28 pixels, and these designs were based on transmissive diffractive layers, each containing ≤ 200×200 trainable phase-only features. To further enhance the unidirectional imaging performance of these diffractive designs, one strategy would be to create deeper architectures with more diffractive layers, also increasing the total number (*N*) of trainable features. In general, deeper diffractive architectures present advantages in terms of their learning speed, output power efficiency, transformation accuracy and spectral multiplexing capability[39,44,47,48]. Suppose an increase in the space-bandwidth product (SBP) of the input FOV A ($SBP_A$) and the output FOV B ($SBP_B$) of the unidirectional imager is desired, for example, due to a larger input FOV and/or an improved resolution demand; in that case, this will necessitate an increase in *N* proportional to $SBP_A × SBP_B$, demanding larger degrees of freedom in the diffractive unidirectional imager to maintain the asymmetric optical mode processing over a larger number of input and output pixels. Similarly, the inclusion of additional diffractive layers and features to be jointly optimized would also be beneficial for processing more complex input spectra through diffractive unidirectional imagers. In addition to the wavelength-multiplexed unidirectional imager reported in Figs. 8-10, an enhanced spectral processing capability through a deeper diffractive architecture may permit unidirectional imaging with e.g., a continuum of wavelengths or a set of discrete wavelength across a desired spectral band. Furthermore, by properly adjusting the diffractive layers and the learnable phase features on each layer, our designs can be adapted to input and output FOVs that have different numbers and/or sizes of pixels, enabling the design of unidirectional imagers with a desired magnification or demagnification factor.

Although the presented diffractive unidirectional imagers are based on spatially coherent illumination, they can also be extended to spatially incoherent input fields by following the same design principles and deep learning-based optimization methods presented in this work. Spatially incoherent input radiation can be processed using phase-only diffractive layers optimized through the same loss functions that we used to design unidirectional imagers reported in our Results. For example, each point of the wavefront of an incoherent field can be decomposed, point by point, into a spherical secondary wave, which coherently propagates through the diffractive phase-only layers; the output intensity pattern will be the superposition of the individual intensity patterns generated by all the secondary waves originating from the input plane, forming the incoherent output image. However, the simulation of the propagation of each incoherent field through the diffractive layers requires a considerably increased number of wave propagation steps compared to the spatially coherent input fields, and as a result, the training of spatially incoherent diffractive imagers would take longer.

## Materials and Methods

**Training details of the diffractive unidirectional imagers.** For the numerical models used in this manuscript, the smallest sampling period for simulating the complex optical fields is set to be identical to the lateral size of the diffractive features, i.e., ~$0.53\lambda$ for $\lambda = 0.75$ mm. The input/output FOVs of these models (i.e., FOV A and B) share the same size of 44.8 mm × 44.8 mm (i.e., ~$59.7\lambda \times 59.7\lambda$) and are discretized into $28 \times 28$ pixels, where an individual pixel corresponds to a size of 1.6 mm (i.e., ~$2.13\lambda$), indicating a $4 \times 4$ binning performed on the simulated optical fields.

For the diffractive model used for the experimental validation of unidirectional imaging, the sampling period of the optical fields and the lateral size of the diffractive features are chosen as 0.24 mm and 0.48 mm, respectively (i.e., $0.32\lambda$ and $0.64\lambda$). This also results in a $2 \times 2$ binning in the sampling space where an individual feature on the diffractive layers corresponds to 4 sampling space pixels that share the same dielectric material thickness value. The input and output FOVs of this model (i.e., FOV A and B) share the same size of 36 mm × 36 mm (i.e., $48\lambda \times 48\lambda$) and are sampled into arrays of $15 \times 15$ pixels, where an individual pixel has a size of 2.4 mm (i.e., $3.2\lambda$), indicating that a $10 \times 10$ binning is performed at the input/output fields in the numerical simulation.

During the training process of our diffractive models, an image augmentation strategy is also adopted to enhance their generalization capabilities. We implemented random translation, random up-to-down and random left-to-right flipping of the input images using the *transforms.RandomAffine* function built-in PyTorch. The translation amount was uniformly sampled within a range of [-10, 10] and [-5, 5] pixels in the diffractive unidirectional imager models used for numerical analysis and the model used for the experimental validation, respectively. The flipping operation is set to be performed at a probability of 0.5.

All the diffractive imager models used in this work were trained using PyTorch (v1.11.0, Meta Platforms Inc.). We selected AdamW optimizer[50,51], and its parameters were taken as the default values and kept identical in each model. The batch size was set as 32. The learning rate, starting from an initial value of 0.03, was set to decay at a rate of 0.5 every 10 epochs, respectively. The training of the diffractive models was performed with 50 epochs. For the training of our diffractive models, we used a workstation with a GeForce GTX 1080Ti graphical processing unit (GPU, Nvidia Inc.) and Core i7 8700 central processing unit (CPU, Intel Inc.) and 64 GB of RAM, running Windows 10 operating system (Microsoft Inc.).

**Vaccination of the diffractive unidirectional imager against experimental misalignments.** During the training of the diffractive unidirectional imager design for experimental validation, possible inaccuracies imposed by the fabrication and/or mechanical assembly processes were taken into account in our numerical model by treating them as random 3D displacements ($D$) applied to the diffractive layers[52]. $D$ can be written as:

$$D = (D_x, D_y, D_z) \qquad (1),$$

where $D_x$ and $D_y$ represent the random lateral displacement of a diffractive layer along the x and y directions, respectively, and $D_z$ represents the random perturbation added to the axial spacing between any two adjacent layers (including diffractive layers, input FOV A and output FOV B). $D_x$, $D_y$ and $D_z$ of each diffractive layer were independently sampled based on the following uniform (U) random distributions:

$$D_x \sim U(-\Delta_{x,tr}, \Delta_{x,tr}) \qquad (2),$$

$$D_y \sim U(-\Delta_{y,tr}, \Delta_{y,tr}) \qquad (3),$$

$$D_z \sim U(-\Delta_{z,tr}, \Delta_{z,tr}) \qquad (4).$$

where $\Delta_{*,tr}$ denotes the maximum amount of shift allowed along the corresponding axis, which was set as $\Delta_{x,tr} = \Delta_{y,tr} = 0.48$ mm (i.e., $0.64\lambda$) and $\Delta_{z,tr} = 1.5$ mm (i.e., $2\lambda$) during the training process. Following the training under this vaccination strategy, the resulting diffractive unidirectional imager shows resilience against possible misalignments in the fabrication and assembly of the diffractive layers.

**Experimental terahertz imaging set-up.** We fabricated the diffractive layers using a 3D printer (PR110, CADworks3D). The test objects were also 3D printed (Objet30 Pro, Stratasys) and coated with aluminum foil to define the light blocking areas, with the remaining openings defining the transmission areas. We used a holder that was also 3D printed (Objet30 Pro, Stratasys) to assemble the printed diffractive layers along with input objects, following the relative positions of these components in our numerical design.

A THz continuous wave scanning system was used for testing our diffractive unidirectional imager design. According to the experimental set-up illustrated in Fig. 6b, we used a THz source in form of a WR2.2 modular amplifier/multiplier chain (AMC) followed by a compatible diagonal horn antenna (Virginia Diode Inc.). A 10 dBm RF input signal at 11.1111 GHz ($f_{RF1}$) at the input of AMC is multiplied 36 times to generate the output radiation at 400 GHz, corresponding to a wavelength of $\lambda = 0.75$ mm. The AMC output was also modulated with a 1 kHz square wave for lock-in detection. The assembled diffractive unidirectional imager is placed ~600 mm away from the exit aperture of the horn antenna, which results in an approximately uniform plane wave impinging on its input FOV (A) with a size of 36 mm × 36 mm (i.e., $48\lambda \times 48\lambda$). The intensity distribution within the output FOV (B) of the diffractive unidirectional imager was scanned at a step size of 1 mm by a single-pixel Mixer/AMC (Virginia Diode Inc.) detector on an XY positioning stage that was built by combining two linear motorized stages (Thorlabs NRT100). The detector also receives a 10 dBm sinusoidal signal at 11.083 GHz ($f_{RF2}$) as a local oscillator for mixing to down-convert the output signal to 1 GHz. The signal is then fed into a low-noise amplifier (Mini-Circuits ZRL-1150-LN+) with a gain of 80 dBm followed by a bandpass filter at 1 GHz (± 10 MHz) (KL Electronics 3C40-1000/T10-O/O), so that the noise components coming from unwanted frequency bands can be mitigated. Then, after passing through a tunable attenuator (HP 8495B) used for linear calibration, the final signal is sent to a low-noise power detector (Mini-Circuits ZX47-60). The detector output voltage is measured by a lock-in amplifier (Stanford Research SR830) with the 1 kHz square wave used as the reference signal. Finally, the lock-in amplifier readings were calibrated into a linear scale. In our postprocessing, linear interpolation was applied to each measurement of the intensity field to match the pixel size of the output FOV (B) used in the design phase, resulting in the output measurement images shown in Fig. 7c.

## Supplementary Materials:

This file contains the *Supplementary Materials & Methods* section and *Supplementary Figures S1-S3*. The Supplementary Materials & Methods includes:

- Numerical forward model of a diffractive unidirectional imager
- Training loss functions and image quantification metrics

# Figures

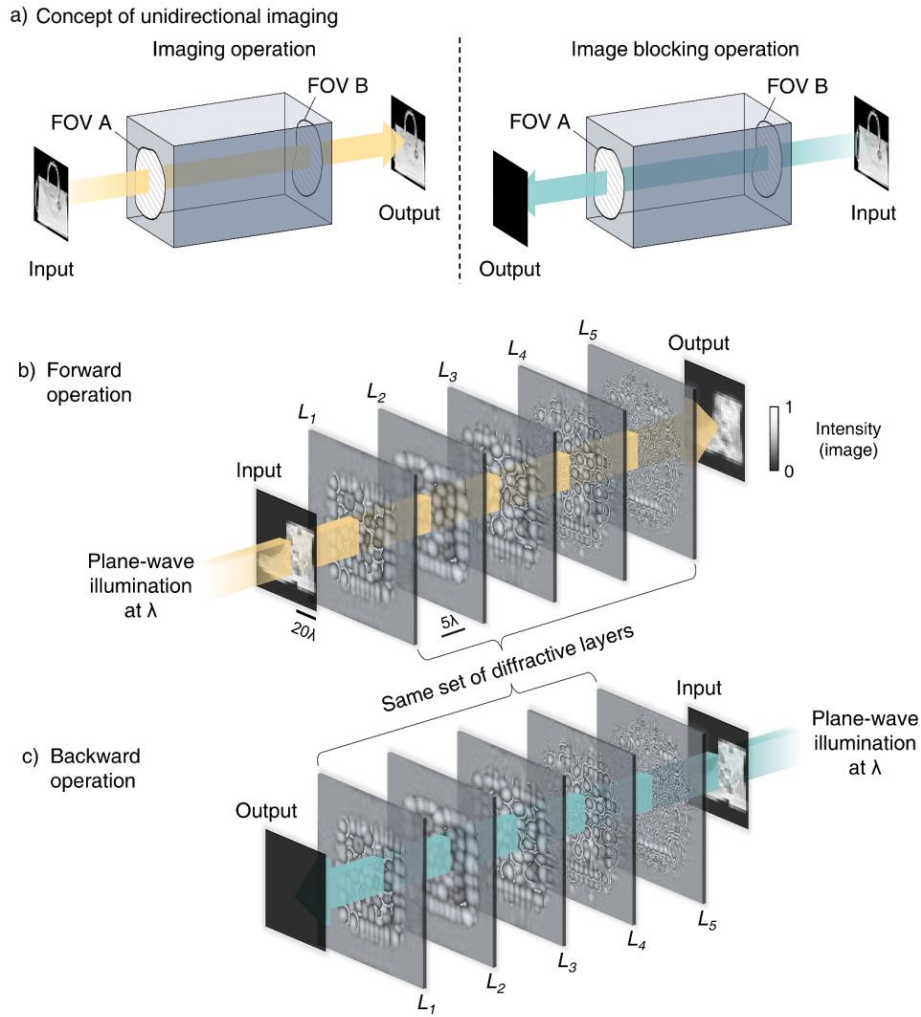

**Figure 1. Schematic of a diffractive unidirectional imager. a,** Concept of unidirectional imaging, where the imaging operation can be performed as the light passes along a certain specified direction (A→B), while the image formation is blocked along the opposite direction (B→A). **b and c,** Illustration of our diffractive unidirectional imager, which performs imaging of the input FOV with high fidelity in its forward (b) direction and blocks the image formation in its backward (c) direction. This diffractive unidirectional imager is a reciprocal device that is linear and time-invariant, and provides asymmetric optical mode processing in the forward and backward directions. Its design is insensitive to light polarization, and leaves the input polarization state unchanged at its output. Furthermore, it maintains its unidirectional imaging functionality over a large spectral band and works under broadband illumination.

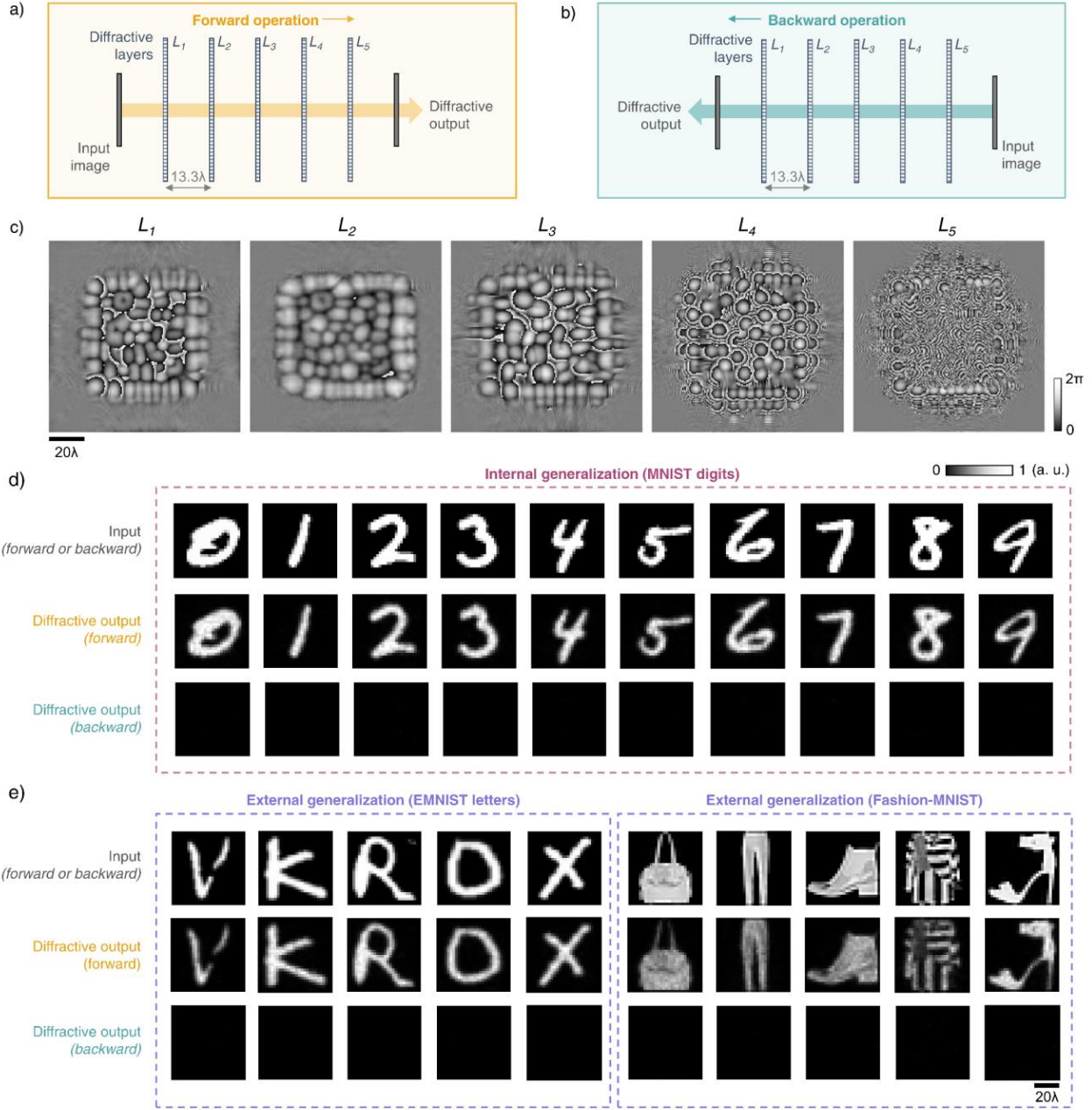

**Figure 2. Design schematic and blind testing results of the diffractive unidirectional imager. a, b,** Layout of the diffractive unidirectional imager when it operates in the forward (a) and backward (b) directions. **c,** The resulting diffractive layers of a diffractive unidirectional imager. **d,** Exemplary blind testing input images taken from MNIST handwritten digits that were never seen by the diffractive imager model during its training, along with their corresponding diffractive output images in the forward and backward directions. **e,** Same as (d), except that the testing images are taken from the EMNIST and Fashion-MNIST datasets, demonstrating external generalization to more complicated image datasets.

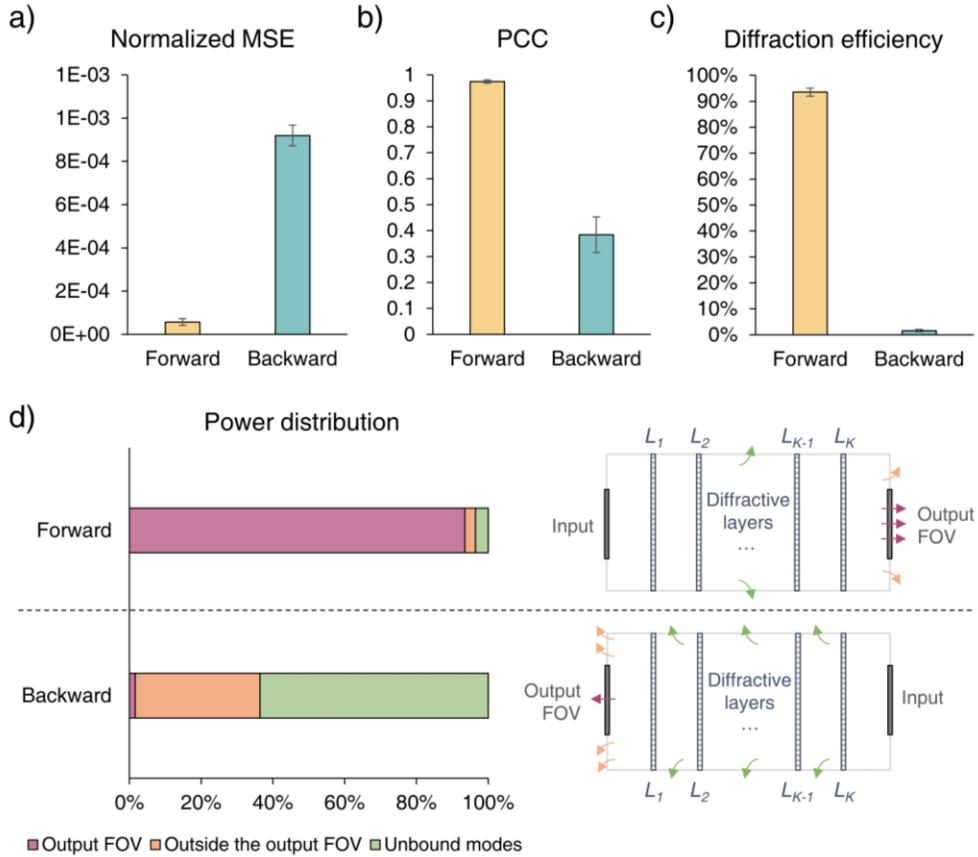

**Figure 3. Performance analysis of the diffractive unidirectional imager shown in Fig. 2. a and b,** Normalized MSE (a) and PCC (b) values calculated between the input images and their corresponding diffractive outputs in the forward and backward directions. **c,** The output diffraction efficiencies of the diffractive unidirectional imager calculated in the forward and backward directions. In (a)-(c), the metrics are benchmarked across the entire MNIST test dataset and reported here with their mean values and standard deviations added as error bars. **d,** (Left) The power of the different spatial modes propagating in the diffractive volume during the forward and backward operations, shown as percentages of the total input power. (Right) Schematic of the different spatial modes propagating in the diffractive volume.
19

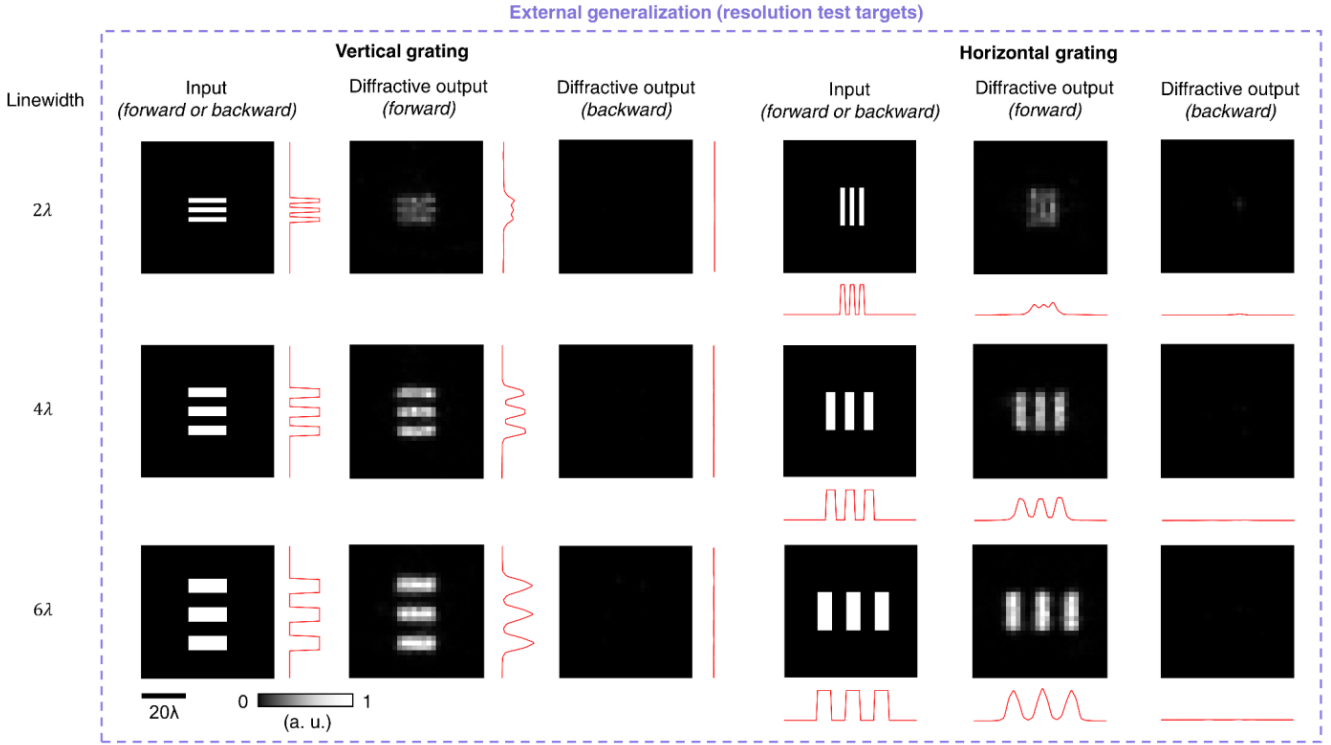

**Figure 4. Spatial resolution analysis for the diffractive unidirectional imager shown in Fig. 2.** Resolution test target images composed of grating patterns with different periods and orientations and their corresponding diffractive output images are shown for both the forward and backward imaging directions. The red lines indicate the one-dimensional cross-sectional profiles calculated by integrating the intensity of the grating patterns in the diffractive output images along the direction perpendicular to the grating.

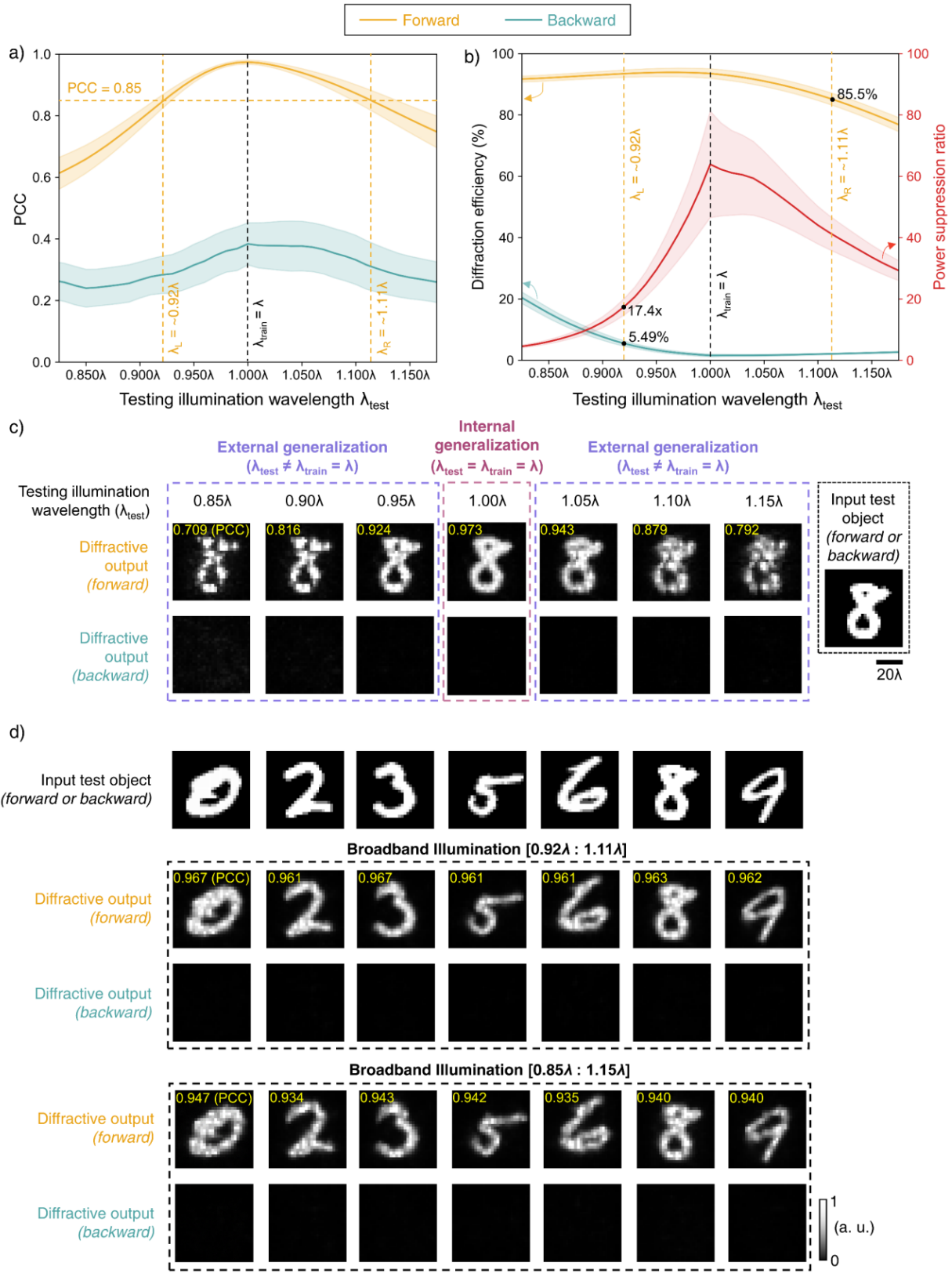

**Figure 5. Spectral response of the diffractive unidirectional imager design shown in Fig. 2. a and b,** Output image PCC (a) and diffraction efficiency (b) of the diffractive unidirectional imager

in the forward and backward directions as a function of the illumination wavelength used during the blind testing. The values of the power suppression ratio are also reported in (b) which refers to the ratio between the output diffraction efficiency of the forward operation and the backward operation. The shaded areas indicate the standard deviation values calculated based on all the 10,000 images in the testing dataset. **c,** Examples of the output images in the forward and backward directions when using different illumination wavelengths during the testing, along with the corresponding input test images (never used during the training). **d,** Broadband illumination results for several test objects are shown for the forward and backward imaging directions. Two different broadband illumination cases are shown, uniformly covering (i) 0.92×λ to 1.11×λ and (ii) 0.85×λ to 1.15×λ, where λ is the training illumination wavelength, $\lambda_{train}=\lambda$.

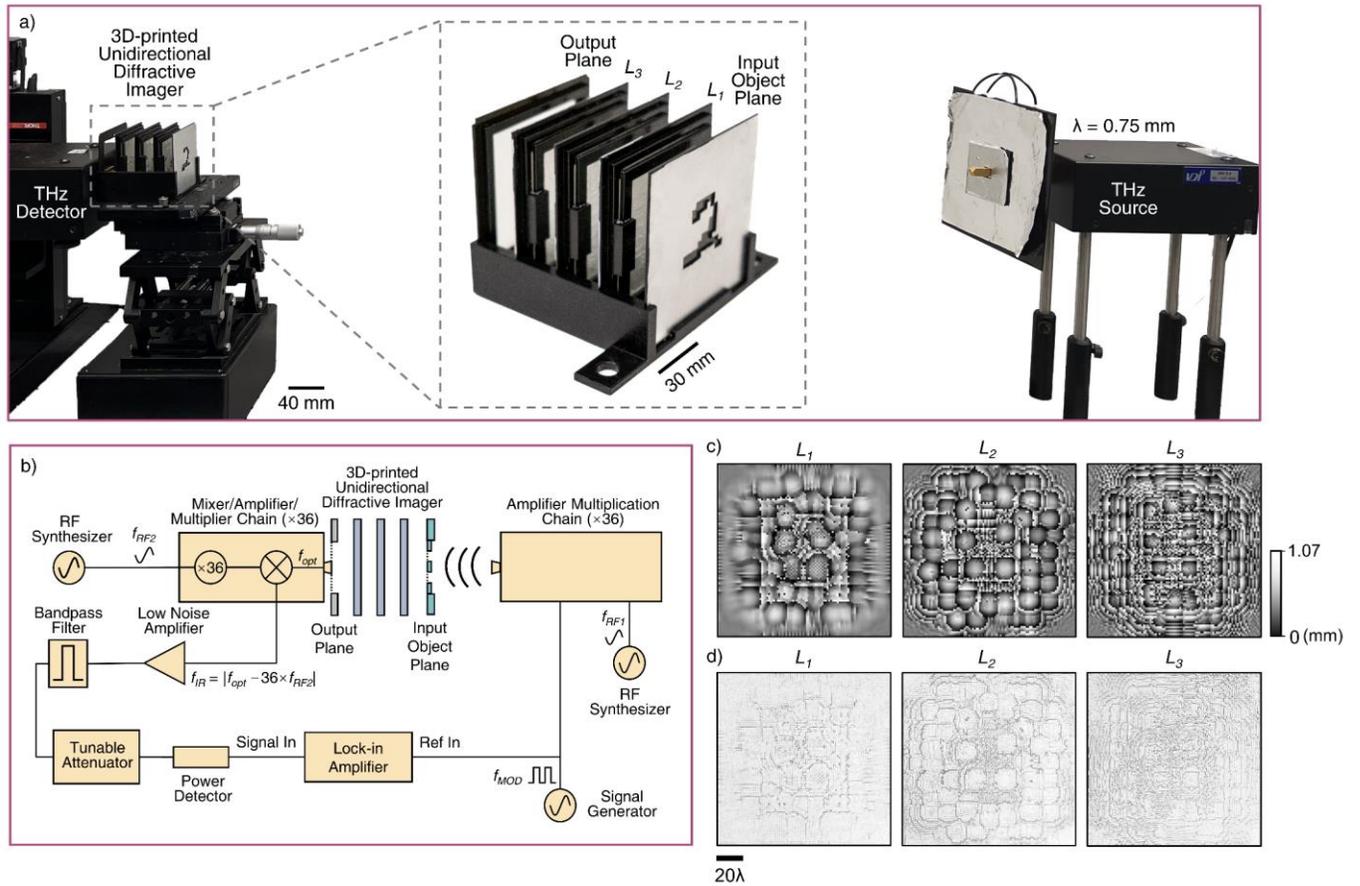

**Figure 6. Experimental set-up for the diffractive unidirectional imager. a,** Photograph of the experimental set-up, including the fabricated diffractive unidirectional imager. **b,** Schematic diagram of the continuous-wave terahertz imaging set-up. **c,** The learned phase profiles of the diffractive layers ($L_1$, $L_2$ and $L_3$). **d,** Photographs of the 3D printed diffractive layers.

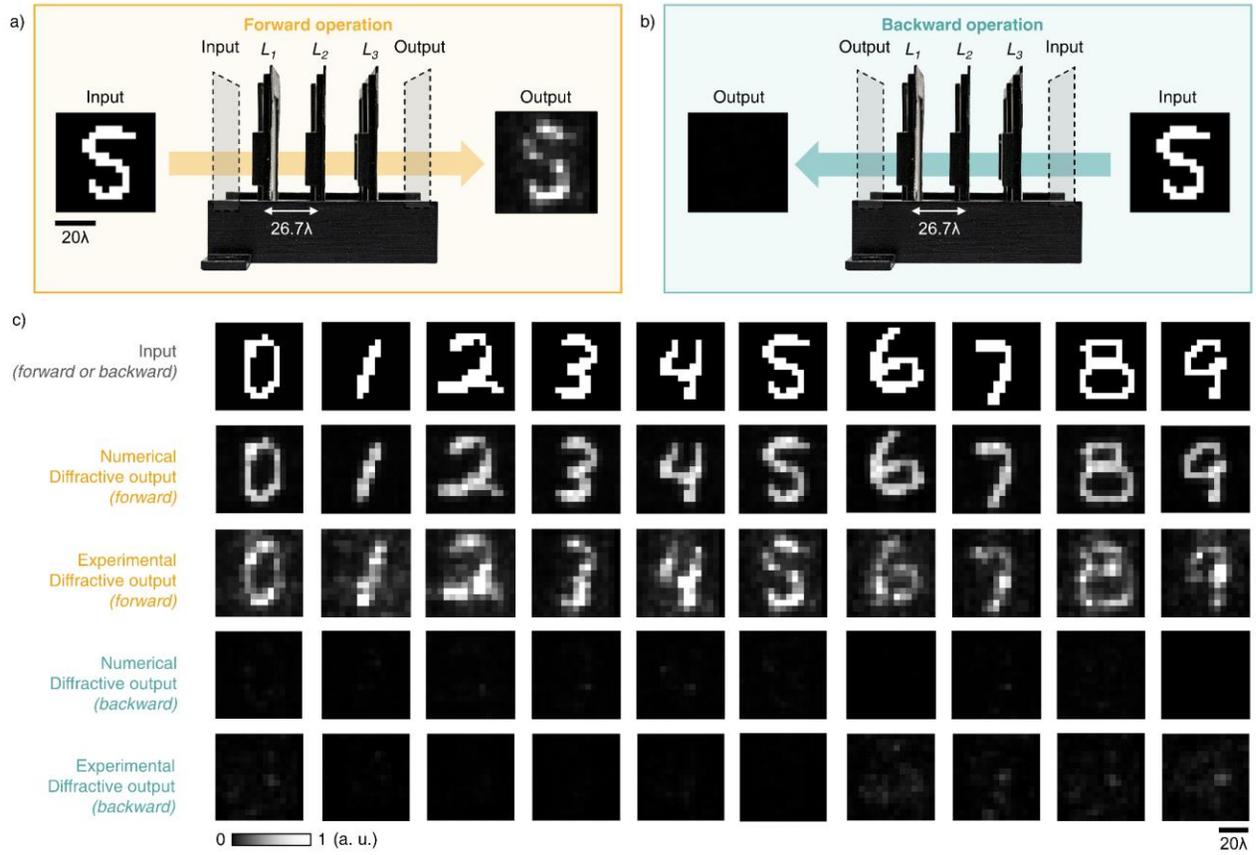

**Figure 7. Experimental results. a-b,** Layout of the diffractive unidirectional imager that was fabricated for experimental validation when it operates in the forward (a) and backward (b) directions. **c,** Experimental results of the unidirectional imager using the fabricated diffractive layers.

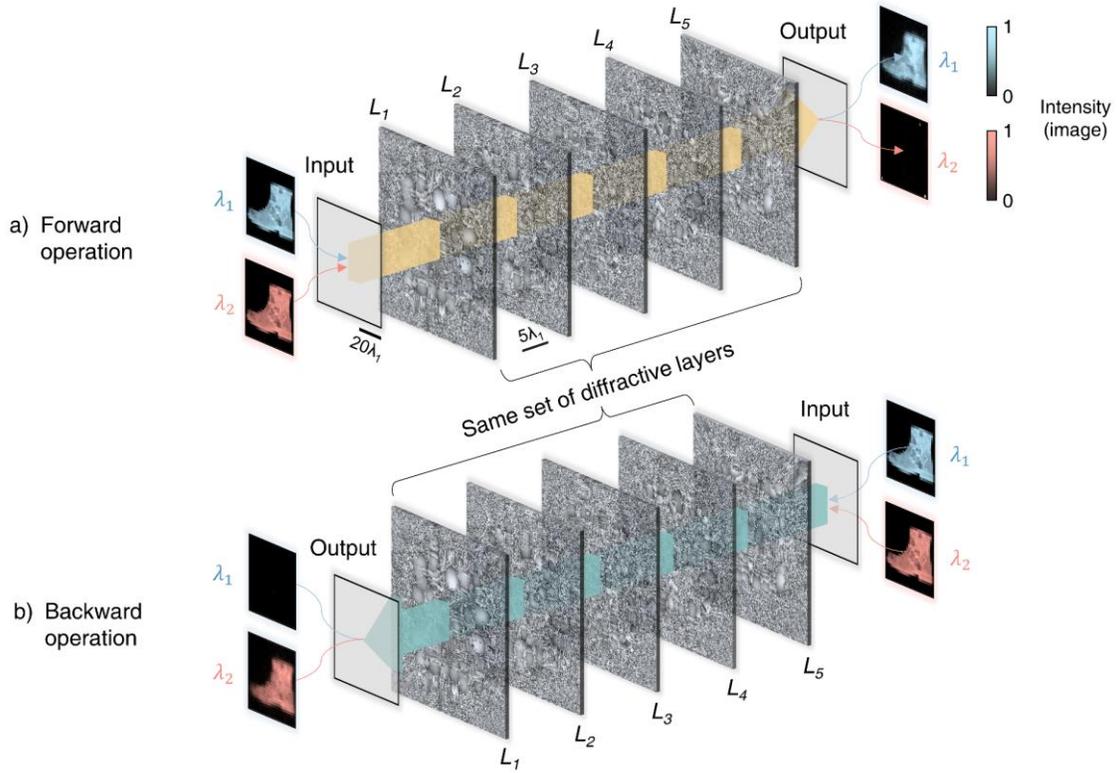

**Figure 8. Illustration of the wavelength-multiplexed unidirectional diffractive imager.** In this diffractive design, the image formation operation is performed along the forward direction at wavelength $\lambda_1$ and the backward direction at $\lambda_2$, while the image blocking operation is performed along the backward direction at $\lambda_1$ and the forward direction at $\lambda_2$. This diffractive imager works as a unidirectional imaging system at two different wavelengths, each with a reverse imaging direction with respect to the other. $\lambda_2 = 1.13 \times \lambda_1$.

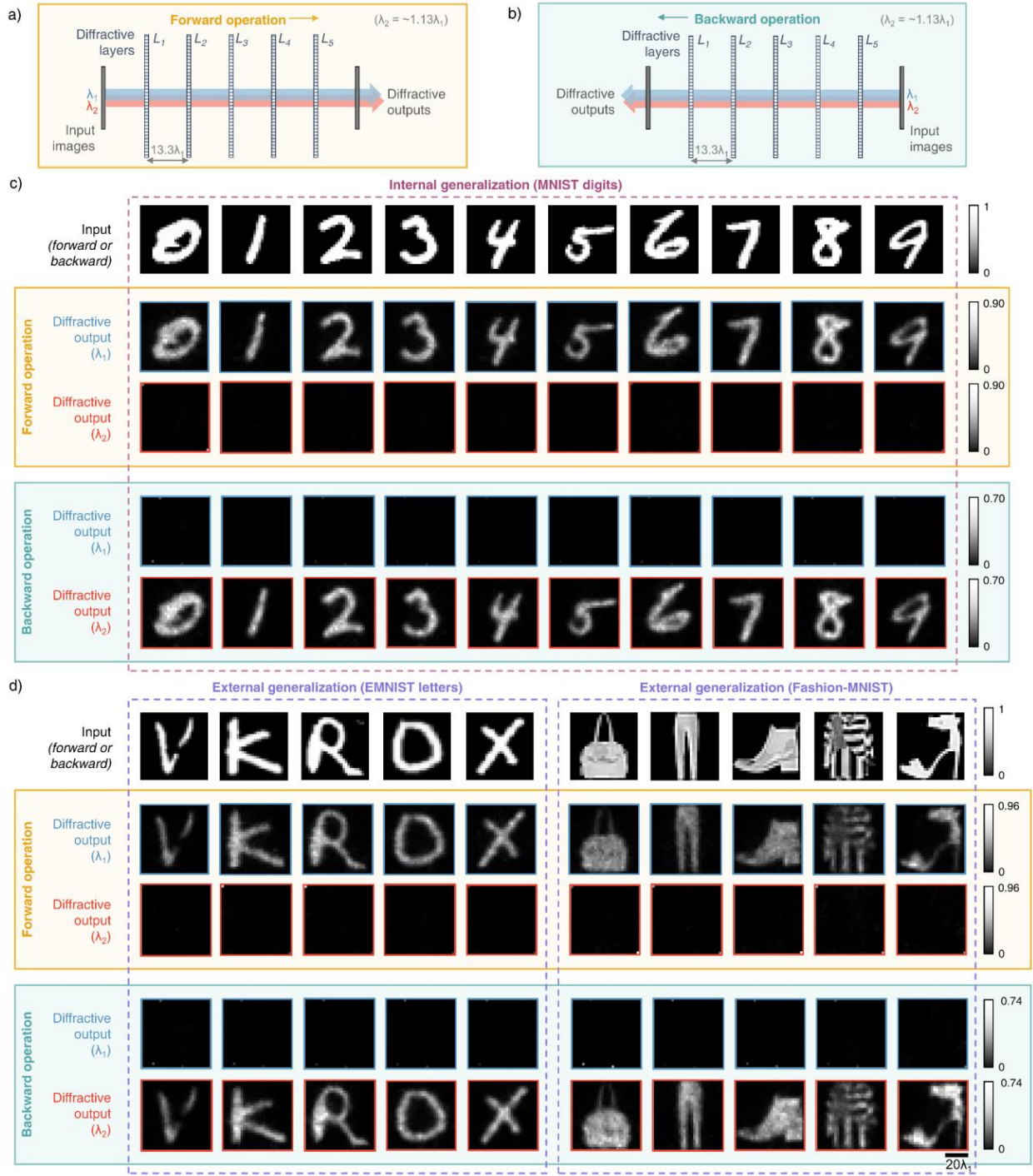

**Figure 9. Design schematic and blind testing results of the wavelength-multiplexed unidirectional diffractive imager. a and b,** Layout of the wavelength-multiplexed unidirectional diffractive imager when it operates in the forward (a) and backward (b) directions. **c,** Exemplary blind testing input images taken from MNIST handwritten digits that were never seen by the diffractive imager model during its training, along with their corresponding diffractive output images at different wavelengths in the forward and backward directions. **d,** Same as (c) except that the testing images are taken from the EMNIST and Fashion-MNIST datasets, demonstrating external generalization.

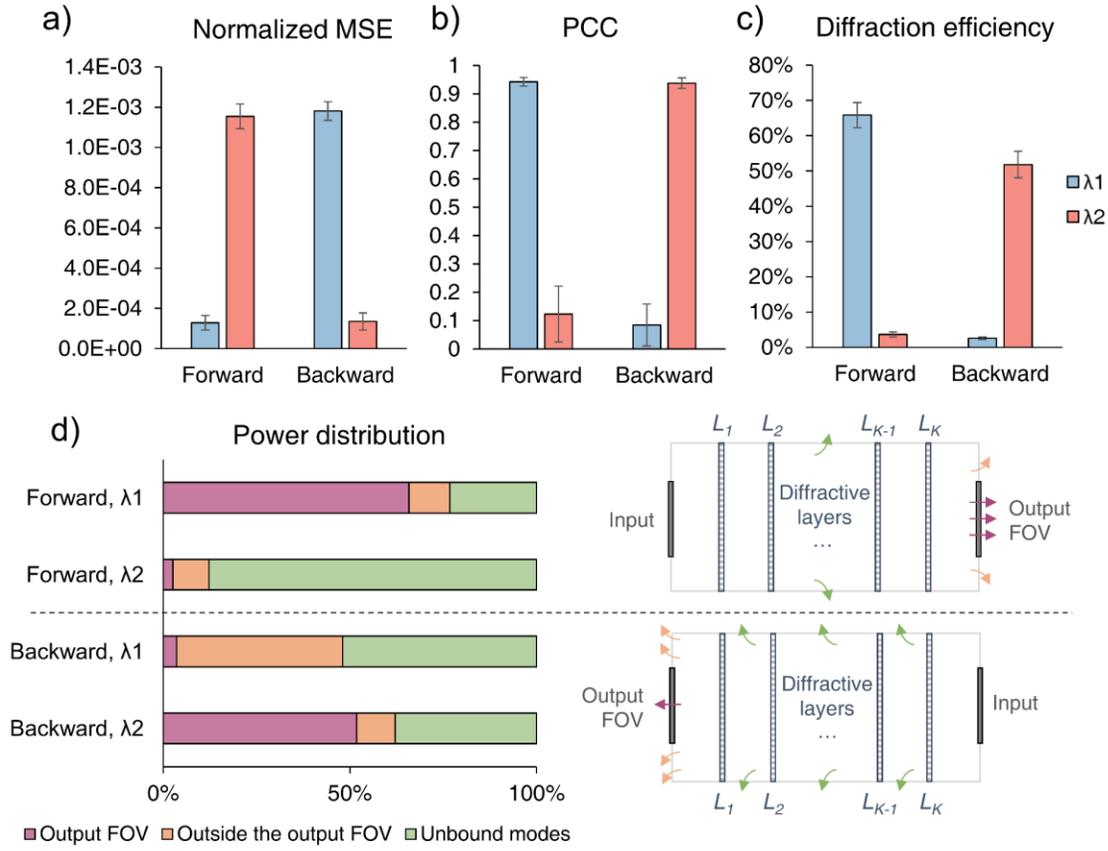

**Figure 10. Performance analysis of the wavelength-multiplexed unidirectional diffractive imager shown in Fig. 9 and Supplementary Fig. S2. a and b,** Normalized MSE (a) and PCC (b) values calculated between the input images and their corresponding diffractive outputs at different wavelengths in the forward and backward operations. **c,** The output diffraction efficiencies of the diffractive imager calculated in the forward and backward operations. In (a)-(c), the metrics are benchmarked across the entire MNIST test dataset and shown with their mean values and standard deviations added as error bars. **d,** (Left) The power of the different spatial modes at the two wavelengths propagating in the diffractive volume during the forward and backward operations, shown as percentages of the total input power (Right) Schematic of the different spatial modes propagating in the diffractive volume.